\documentclass[conference]{IEEEtran}
\usepackage{cite}

 \usepackage{amsbsy}

\ifCLASSINFOpdf
\else
\fi

\usepackage{graphicx}
\usepackage{url}

\begin{document}
%
\title{Breast Cancer Data Analytics With Missing Values: A study on Ethnic, Age and Income Groups}

\author{\IEEEauthorblockN{Santosh Tirunagari\IEEEauthorrefmark{1},
Norman Poh\IEEEauthorrefmark{2}, Hajara Abdulrahman\IEEEauthorrefmark{3}, Nawal Nemmour\IEEEauthorrefmark{4}
and
David Windridge\IEEEauthorrefmark{5}}
\\\IEEEauthorblockA{University of Surrey, Guildford, Surrey GU2 7XH\\
\IEEEauthorrefmark{1}s.tirunagari@surrey.ac.uk,
\IEEEauthorrefmark{2}n.poh@surrey.ac.uk,
\IEEEauthorrefmark{3}ha00162@surrey.ac.uk,
\\\IEEEauthorrefmark{4}nn00070@surrey.ac.uk,
\IEEEauthorrefmark{5}d.windridge@surrey.ac.uk}}


\maketitle

\begin{abstract}
An analysis of breast cancer incidences in women
and the relationship between ethnicity and survival rate has been
an ongoing study with recorded incidences of missing values in the
secondary data. In this paper, we study and report the
results of breast cancer survival rate by ethnicity, age and income groups from the dataset collected for 53593 patients in South East England between the years 1998 and 2003. 
In addition to this, we also predict the missing values for the ethnic groups in the dataset. The principle findings in our study suggest that: 1) women of white ethnicity in South
East England have a highest percentage of survival rate when compared to the black ethnicity, 2) High income groups have higher survival rates to that of lower income groups and 3) Age groups between 80-95
and 20-35 have lower percentage of survival rate.

\end{abstract}


%
\IEEEpeerreviewmaketitle

\section{Introduction}
\label{motiv}
\subsection{Motivation for imputation}

Surveys often contain missing values and the process of replacing these missing values with substituted values is called as imputation~\cite{rubin2004multiple}~\cite{brick1996handling}. These missing values might induce more useful information in predicting the trends and statistics. For conveniences of analysis, statisticians normally discard observations containing the missing values. This results in the reduction of the sample size for the analysis and interesting information might get lost. This can be a serious hindrance not only
by misleading the results~\cite{schafer1999multiple} but also by producing overly simplified conclusions. Hence, there is a need for imputation when dealing with the data analysis especially in the health care domain. 

From the literature the missing value problem can be categorised into two main types: 1) missing values at random and 2) missing values not at random~\cite{little2002statistical}. Missing values not at random can again be of two types 1) for discrete values (group membership)~\cite{tabachnick2001using} and 2) for continues values. Methods such as: 1) mean substitution, 2) median substitution could be utilises for computing the missing values at random~\cite{allison2002missing}~\cite{little2002statistical}. Other method including, maximum votes and nearest neighbours techniques can also be employed for this purpose. Regression based substitution can be employed for continuous missing values that are not at random. The aforementioned methods fail when imputing the values for group membership (refer to~\cite{sethi1993optimism} for computing missing values for group memberships). When dealing with missing values for group memberships, methods such as: 1) analysis of variance (AnoVa)~\cite{hoaglin1978hat}, 2) analysis of mean and 3) classification based methods~\cite{lowd2005naive} could be employed. In the earlier methods, the variance and mean for each group is calculated and mean square error is computed between the observations of group and the new group that contains missing values. The group member with minimum mean square error is assigned to the missing value. In the classification based approach, the missing values are considered as the unknown class labels and a classifier model is built on the existing observations. Using the classifier model the missing values are then computed. In this process, the classifier must be cross validated for the performance.

Using traditional statistical methods sometimes might give us overly simplified solutions, hence there is a need for statistical machine learning algorithms which are a combination of both probability and statistics~\cite{lowd2005naive}. Therefore, we in this study use classification based method to impute the missing values for group membership, i.e ethnicity group. 

\subsection{Motivation underlying the current study}
Breast cancer is a malignant tumour that originates from the
cells of the breast and grows into surrounding and distant tissues~\cite{griffiths2012triple}. It is the second most common cancer in women
which makes this study important. The relationship between
ethnicity, survival percentage, and breast cancer is complex.
Studies carried out in the past have shown that women of
different ethnicity have different rates of survival from breast
cancer after diagnosis~\cite{eley1994racial}. Many comparisons and links have been made between
ethnicity, income and survival rates in women diagnosed with breast
cancer~\cite{grann2006regional}~\cite{parker1998cancer}~\cite{bradley2002race}. 
According to a study carried out by the Cancer
Research UK~\footnote{\url{http://www.cancerresearchuk.org/}}, after grouping women into ethnicity
groups aged
between 15 and 64 years, the percentage of survival from
breast cancer of those of white ethnicity is relatively higher at
91.4\% than women of black ethnicity with survival percentage
of 85.0\%. The National Cancer Intelligence
Network~\footnote{\url{http://www.ncin.org.uk/}} have produced a report~\cite{rule} on
`Cancer Incidence and Survival by Major Ethnic Groups in
England between 2002 and 2006'. This report shows that the survival rates of women with breast
cancer categories into four major ethnicity groups namely: White, Asian, Black and Unknown, white women had a higher rate of survival compared to those of black ethnicity. Bradley $et$ $al.$ in~\cite{bradley2002race}, showed that low socioeconomic status was associated with late-stage breast cancer at diagnosis and mostly in death.

\subsection{Objectives and contribution}
\label{obj}

The main objectives of this study are two-fold. On one hand, from the computational perspective, we would like to examine the feasibility of using machine learning based classifier (e.g, Naive Bayes) in filling up the missing values in the data. On the other hand, from the scientific perspective, we would like understand whether the insights from breast cancer analytics correspond to what clinicians would expect. For example to answer the following questions that are exploratory in nature.

\begin{itemize}
\item Does survival rate get affected by the age and ethnic group
of the patient? for instance, are black and older women more
likely to die if they have breast cancer.
\item Does financial status of the patient have any effect on the survival rate? i.e, do wealthier have
 lower possibility of dying from the breast cancer.
\end{itemize}

Our contribution in this paper is to show how a machine learning based classifier can be utilised to impute the missing values in the health care data and obtain insights. 
When there are many classification methods available in the literature, it is difficult to choose which one to use. In such a case simplicity, reputation of the method and experience of its usage can influence the selection process. Therefore, in this study, we have chosen Naive Bayes classifier to compute the missing values because of its simplicity and inexpensiveness.

\subsection{Organisation}
 The organisation of the paper is as follows: In section~\ref{data}, we present and analyse the breast cancer dataset. In section~\ref{methods}, we discuss our methodology. Results are discussed in section~\ref{results}. Finally, in section~\ref{condis}, we draw conclusions and summarise with the discussions.

\section{Data preprocessing}
\label{data}

The dataset is collected for 53593 breast cancer incidences in
women taken in South East England between the years 1998 and 2003. 
The initial dataset consists of 13 features however some of these
features are simply an alternative way of representing
the existing ones. Therefore these features are removed from the
dataset. Features such as 'Year of
Diagnosis' and 'Year of Death or Censored' are removed as
this data was available to us within the 'Survival' feature. We
also have removed the single year 'Age at diagnosis' feature as we already have this information within the 'Age' feature. 
This left us with a final dataset of 9 features as shown
in Table~\ref{format} along with its Data format.

\begin{table}[h]
\centering
\caption{Table showing the features and their format.}
\label{format}
\begin{tabular}{|c|c|}
\hline
\textbf{Feature}       & \textbf{Data Format}                       \\ \hline
Income Quintile        & 1 = (Most Affluent) to 5 = (Most Deprived) \\ \hline
Age at Diagnosis Group & 0 = (0-4); 5 = (5-9) to 100 = (100+)       \\ \hline
Ethnic Group           & Ethnic Groups (Table 2)                    \\ \hline
Radiotherapy           & 0=No; 1=Yes                                \\ \hline
Chemo Therapy          & 0=No; 1=Yes                                \\ \hline
Hormone Therapy        & 0=No; 1=Yes                                \\ \hline
Cancer Surgery         & 0=No; 1=Yes                                \\ \hline
Survival days          & Total no. of days                          \\ \hline
Death of Breast Cancer & 0=No; 1=Yes                                \\ \hline
\end{tabular}
\end{table}

The second stage of our data preprocessing is to convert the
ethnic group data from nominal to indices (Table~\ref{nominal}) and
remove header labels from the dataset so that it can be used for further data analysis.

\begin{table}[h]
\centering
\caption{Nominal values of ethnic groups and their corresponding numerical value.}
\label{nominal}
\begin{tabular}{|c|c|c|}
\hline
\textbf{Ethnicity Group} & \textbf{Nominal Values} & \textbf{Indices} \\ \hline
White                    & W                       & 1                \\ \hline
Not Known                & NK                      & 2                \\ \hline
Any Other                & Oth                     & 3                \\ \hline
Black Caribbean          & BC                      & 4                \\ \hline
Chinese                  & C                       & 5                \\ \hline
Indian                   & In                      & 6                \\ \hline
Black African            & BA                      & 7                \\ \hline
Pakistani                & P                       & 8                \\ \hline
Black Other              & BO                      & 9                \\ \hline
Asian Other              & AO                      & 10               \\ \hline
Mixed                    & M                       & 11               \\ \hline
Bangladeshi              & Ba                      & 12               \\ \hline
\end{tabular}
\end{table}

\subsection{Demographics}

\begin{figure}[ht]
\centering
\includegraphics[width=3.5in]{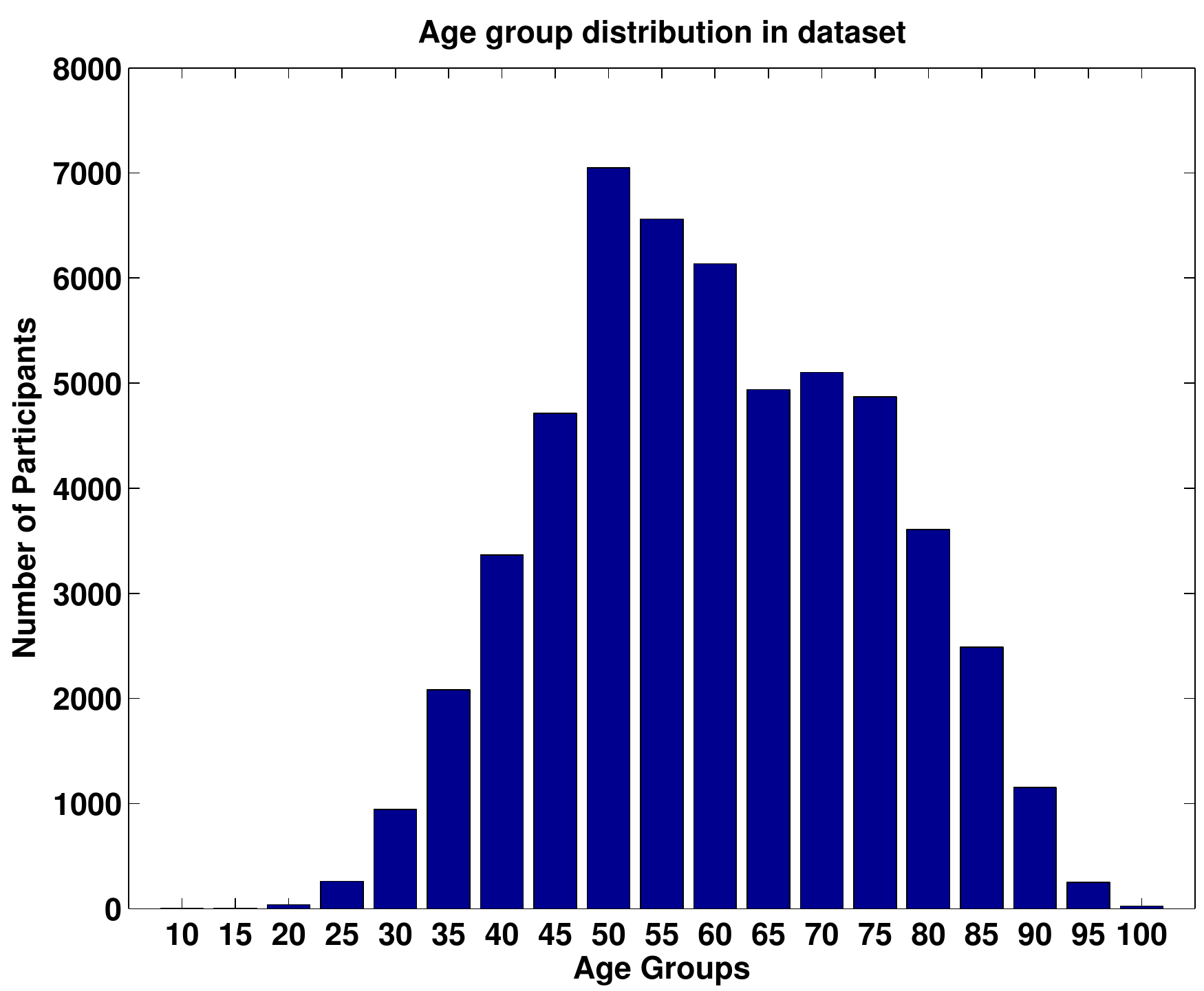}
\caption{Distribution of age within the patients.}
\label{fig_age}
\end{figure}

The age group distribution in the dataset as seen in Figure~\ref{fig_age} shows an expected normal distribution of age within the patients and indicates the highest frequency of them are between 50 and 65 years of age.

\begin{figure}[ht]
\centering
\includegraphics[width=3.5in]{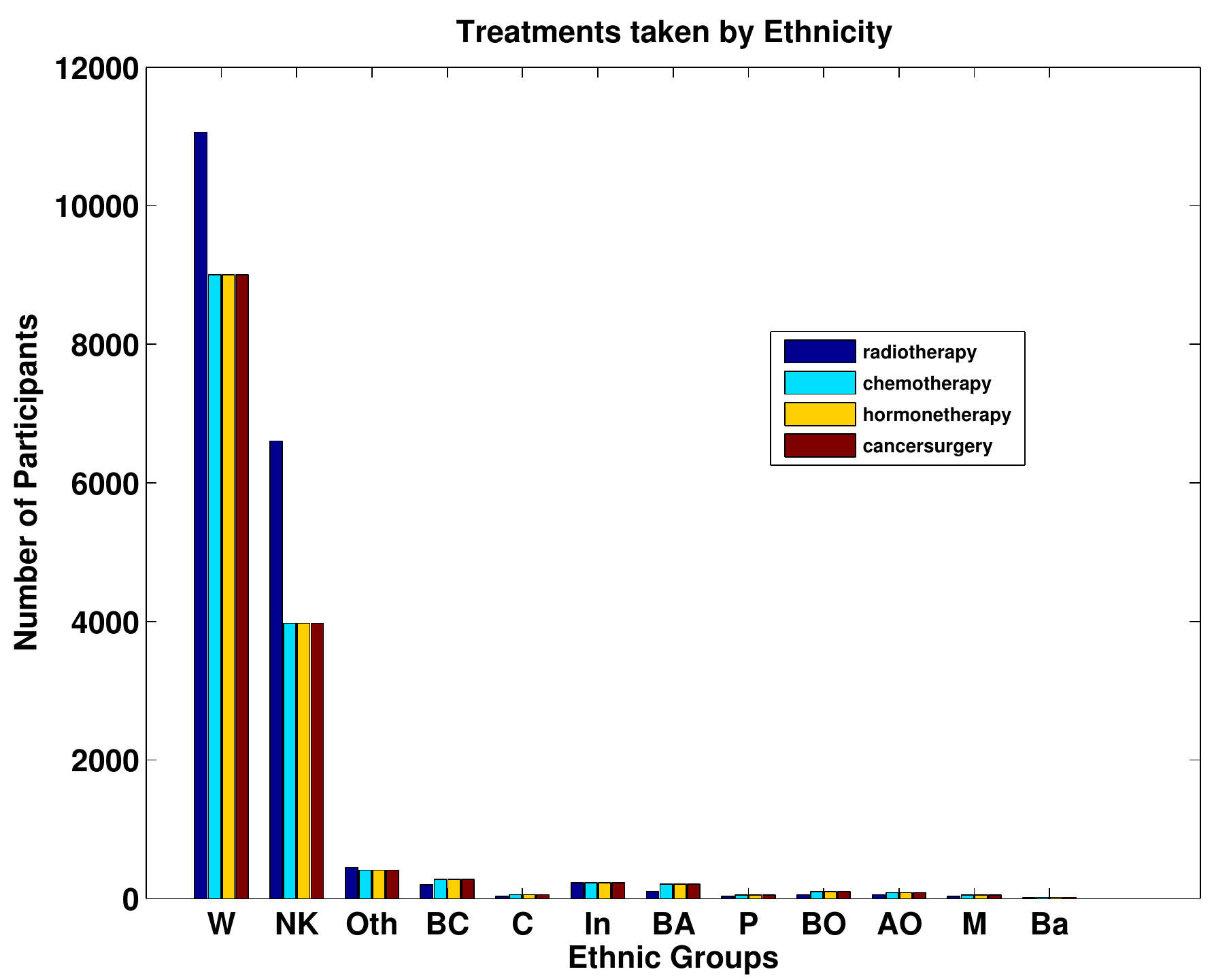}
\caption{Different treatments for breast cancer.}
\label{fig_tre}
\end{figure}

The most common cancer treatment taken by women of white ethnicity is radiotherapy as seen in Figure~\ref{fig_tre}. However this seems to be the least common treatment for women of black African ethnicity. 

\begin{figure}[ht]
\centering
\includegraphics[width=3.5in]{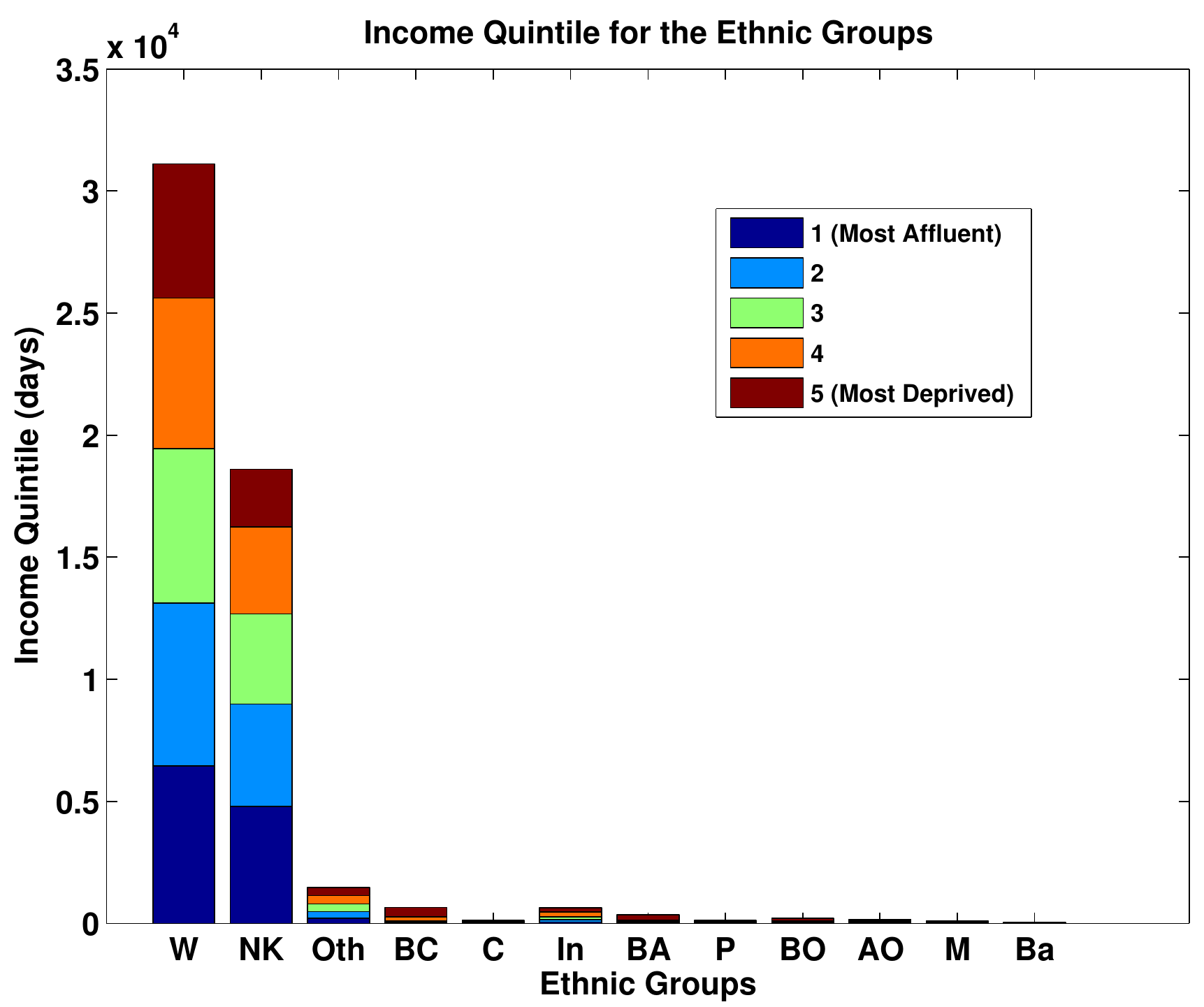}
\caption{Distribution of income within the patients.}
\label{fig_income}
\end{figure}

Figure~\ref{fig_income} shows that the women from the white ethnic group are quite evenly distributed in terms of their socioeconomic deprivation.

\begin{figure}[ht]
\centering
\includegraphics[width=3.5in]{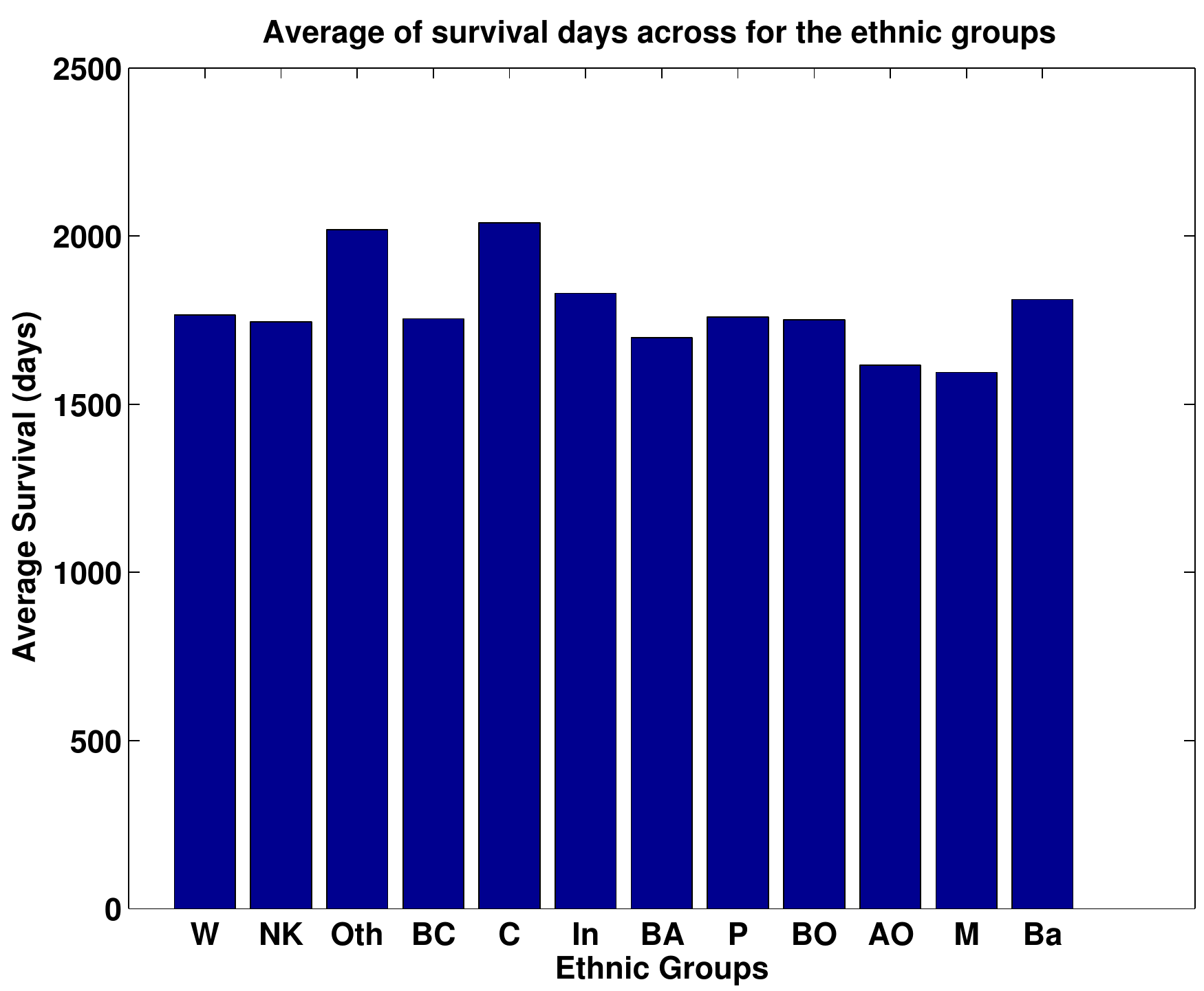}
\caption{Average survival days across the ethnic groups.}
\label{fig_surv}
\end{figure}

Looking at the average of survival days across the ethnic groups shows that women of Chinese ethnicity have the highest days of survival from breast cancer (see, Figure~\ref{fig_surv}). It also highlights that although the proportion of women of white ethnicity is significantly higher than any other groups, an average count gives a better indication of the caner survival rate across the ethnicity. Figure~\ref{fig_surv_et} shows that the proportion of breast cancer survival compared to death is much higher within the white ethnic group and similar in black Caribbean and Indian ethnicity.

\section{Methodology}
\label{methods}
\subsection{Naive Bayes for imputation}
In order to fill up the ethnicity of the records with unknown ethnicity
group, we decided to take a supervised-learning approach. The
supervised-learning (machine learning) uses known set of input
data and response to the data to build a predictor model that
will generate predictions for the response to the new set of data. 
Since our dataset mainly consisted of binary and nominal
data values we determined that this was a classification
problem and chose Naive Bayes classification algorithm to
model our predictor. The Naive Bayes algorithm seemed to be
the optimal choice despite it having a low predictive accuracy
because it handles categorical predictors very well and its
speed and memory usage are good for simple distributions.
More importantly this algorithm is easy to interpret because it
is based on finding the posterior probability for the new data
belonging to the classes given that the features are independent
of one another in each class. 

The Naive Bayes classifier
involves two stages. The first stage is training, where the
probabilities of every features' parameter given each class as
well as the probability of each class are estimated. These are
known as the Likelihood $P(X|C)$ and Class Prior probability
$P(C)$ respectively. The second stage is prediction, where the
posterior probability algorithm (Eq.~\ref{eq:2.8}) calculates the
probability of each class given the parameters of each feature
in the new data. Finally it predicts the class with the highest
posterior probability as the result. As the features of our dataset are
also assumed to be independent of each other and the class we
believe that using the Naive Bayes algorithm will give us the
best output.

\begin{equation}
\label{eq:2.8}
	P(C|X) = \frac{P(X|C) P(C)}{P(X)}.
\end{equation}

\subsection{Parameter tuning}

The Naive Bayes Classifier supports a number of
probability distribution estimates. Based on theory the
Multivariate Multinomial Distribution is the ideal distribution
for us to choose as our dataset consists of categorical features;
however we decided to conduct a set of parameter tuning
experiments with the different distribution options available in Matlab to
observe the legitimacy of the theory. We chose: 1)Normal (Gaussian), 2) Kernel, 3) Multinomial and
4) Multivariate Multinomial and set the class prior probability as
uniform for all cases so that the probabilities are equal for all
classes. subsequently, we decided to choose the
best distribution depending on the highest
value for accuracy after cross validations. 

\subsection{Cross validation}
In order to run the 
cross validation we first extracted the records of unknown
ethnic group from the original dataset and created the training
data with the remaining records. We decided to use the K Fold
Cross validation process in order to enhance the accuracy of
the results and a value of 10 for K seemed ideal for such a large
dataset. The 10 fold cross validation involved dividing our
training data into 10 sets, then setting aside one set for
validation we used the remaining 9 sets to train the Bayesian
classifier. Then we cross validated the results with the
validation set and calculated its accuracy using \emph{unbiased} F-measure~\cite{forman2010apples}. This cross validation was computed 10 times
where every time a different set was used for validation  and then an average of the F-measure
percentages are calculated. After running this for each of the
distribution parameters we chose to use the distribution with
the highest accuracy. We ran the prediction model for the four
distributions we considered.

\subsection{Performance evaluation}
The F-measure is a good way to calculate the performance of a
prediction model by checking the predicted results against the
actual results. The process involves finding the total number of
True Positives (tp), True Negatives (tn), False Positives (fp)
and False Negatives (fn) from the result comparison. Then
finding the Precision and Recall using the equations in Table~\ref{tab:met.5}
where Precision is the ratio of number of correct results to the
number of all returned results and Recall is the ratio of the number of
correct results to the number of results that should
have been returned [11]. Finally the \emph{unbiased} F-measure is then
calculated by finding the harmonic mean of the Precision and
Recall rates. 

\begin{table}[h]
\centering
\caption{Performance measurement methods}
\label{tab:met.5}
\begin{tabular}{ll}
\hline
\textbf{Method} & \textbf{Formula}\\
\hline
Recall & $Re =\frac{TP}{TP+FN}$\\\\
Precision  & $Pr=\frac{TP}{TP+FP}$\\\\
F-measure & $F = 2*\frac{Pr \times Re}{Pr+Re}$\\\hline
\end{tabular} 
\end{table}





\subsection{Imputation}
Once the dataset was finally ready to be classified by the
Bayesian model we assigned the previously extracted records
of the unknown ethnic group as our testing dataset and
keeping all of the remaining data records for training and the
Ethnic group feature was assigned as the class label. The
testing dataset consisted of 18595 records which is around 35\%
of the original records. This actually gives a close 30:70 ratio
between the testing and training which is optimal as previously
mentioned. Once the predicted results were obtained we
integrated the unknown records back into the original dataset
and replaced the unknown values with the predicted ethnic
groups.

\section{Results}
\label{results}

In this section, we present the results for the following:

\begin{itemize}
\item Cross validation results for considered distributions.
\item Show the effect of ethnicity on the survival rate.
\item Show the impact of age on the survival rate.
\item Show the implication between financial status and the survival rate.
\end{itemize}

\subsection{Cross validation}
Table~\ref{fmes} indicates that fitting the Bayesian model with a
'Kernel' distribution with uniform prior and 'Gaussian' distribution with no prior as parameters give the most
accurate 94.10\% and 94.01\% results respectively when compared to the other distributions. Therefore, we consider predictions based on kernel distribution for the imputation.
 
\begin{table}[h]
\centering
\caption{F-measure \% according to each distribution.}
\label{fmes}
\begin{tabular}{|l|l|}
\hline
\textbf{Distribution type}                  & \textbf{F-measure \%} \\ \hline
Gaussian with no prior                      & 94.01\%               \\ \hline
Gaussian with uniform prior                 & 48.7\%                \\ \hline
Kernel with uniform prior                   & 94.10\%               \\ \hline
Multinomial with uniform prior              & 41.98\%               \\ \hline
Multinomial Multivariate with uniform prior & 50.80\%               \\ \hline
\end{tabular}
\end{table}

\subsection{Ethnic groups vs survival rates}
\subsubsection{Before imputation}
\begin{figure}[ht]
\centering
\includegraphics[width=3.5in]{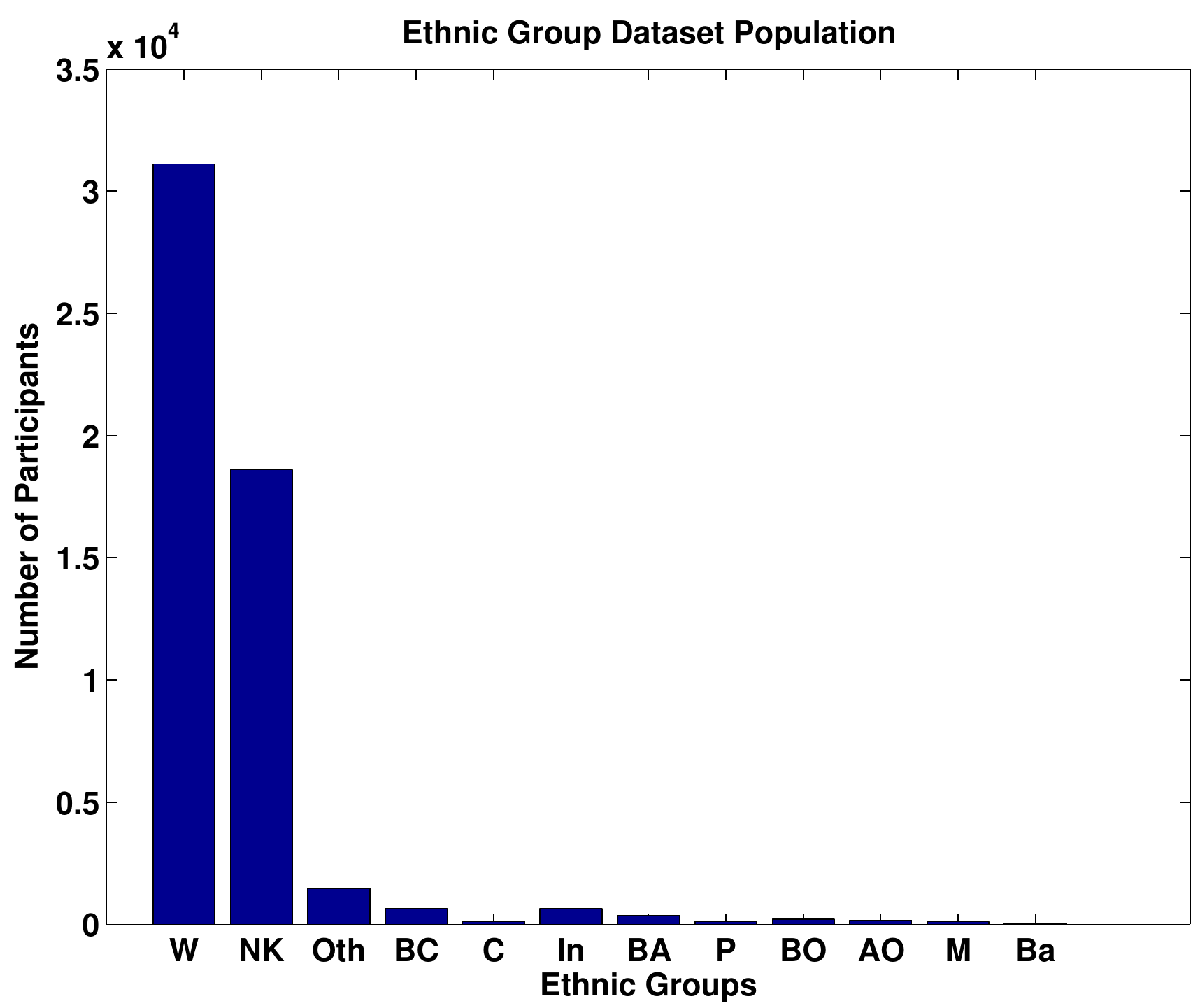}
\caption{Distribution of ethnic groups with the majority of participants being white.}
\label{fig_eth}
\end{figure}

Figure~\ref{fig_eth} shows the distribution of ethnicity within our original dataset. 
According to that, the
majority of our samples are white women (58\%). A high
percentage (35\%) of the samples are from unknown ethnic
groups. Other ethnic groups exist in small percentages with
Bangladesh being the lowest. The high number of white
women is due to socio-demographic reasons. The data was
collected in Southwest of England where most of the
population is of white ethnic group. This is justified by the data produced by the Office for National
Statistics census data, UK~\cite{CODE}. The population in
South East England by ethnic group in 2009 contains 90.7\% of
white ethnicity.

The distribution of data
affects our results due to the unequal number of samples
between the ethnic groups. In
order to avoid that, we convert the existing numbers to percentages so as to make results more reliable. 

The results for the mortality rate
according to ethnicity show that the highest number of people
that died of breast cancer is in white ethnic group (Figure~\ref{fig_eth}).
This does not necessarily mean that this group is more probable
to die from breast cancer. In order to get the possibility
of each ethnic group facing cancer we translate our
results in percentages within each group. 

\begin{figure}[ht]
\centering
\includegraphics[width=3.5in]{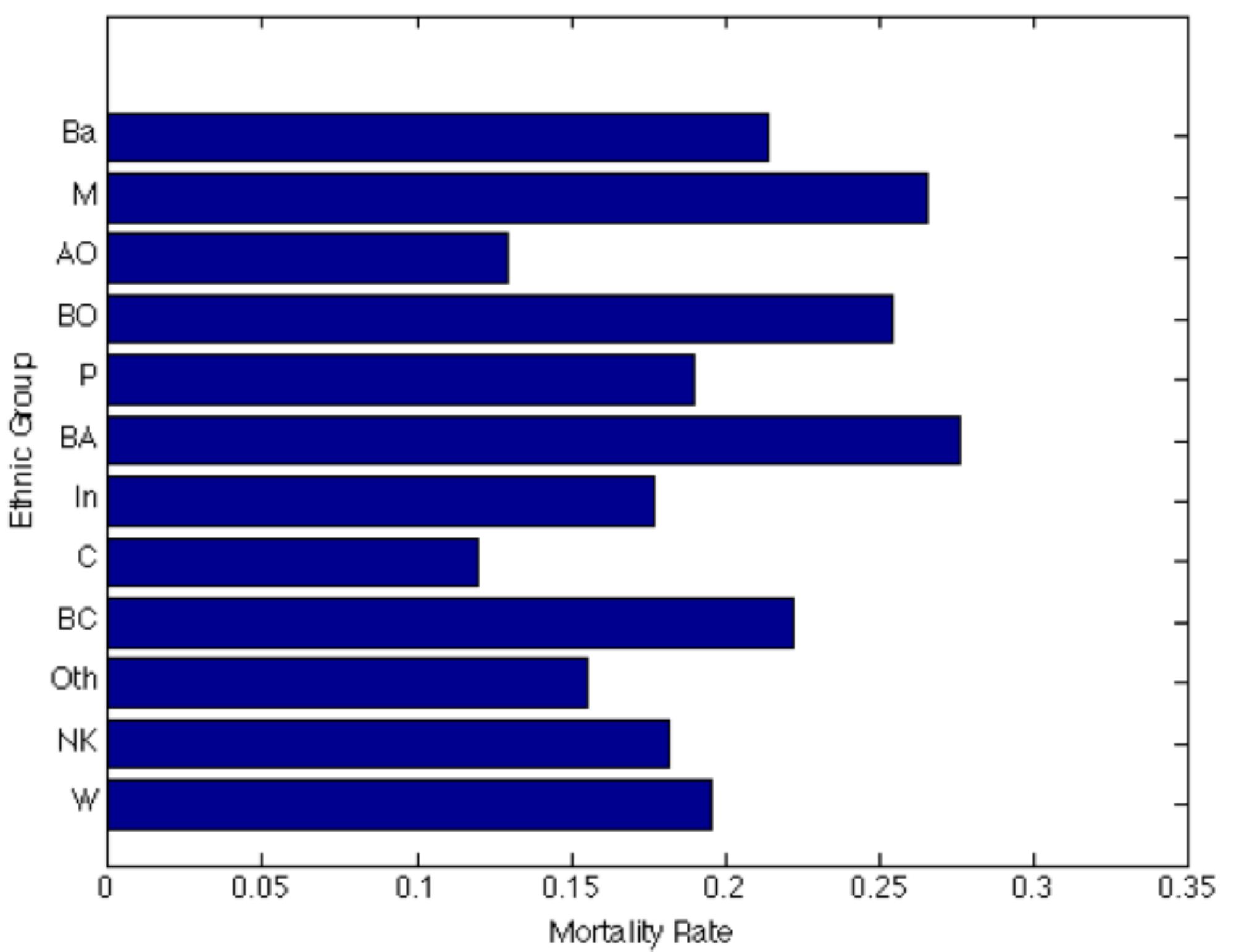}
\caption{Distribution of ethnic groups with the majority of participants being white.}
\label{fig_eth_per}
\end{figure}

Figure~\ref{fig_eth_per} shows that
white women have lower mortality rate than black African,
mixed and black other.

\subsubsection{After imputation}

\begin{figure}[ht]
\centering
\includegraphics[width=3.5in]{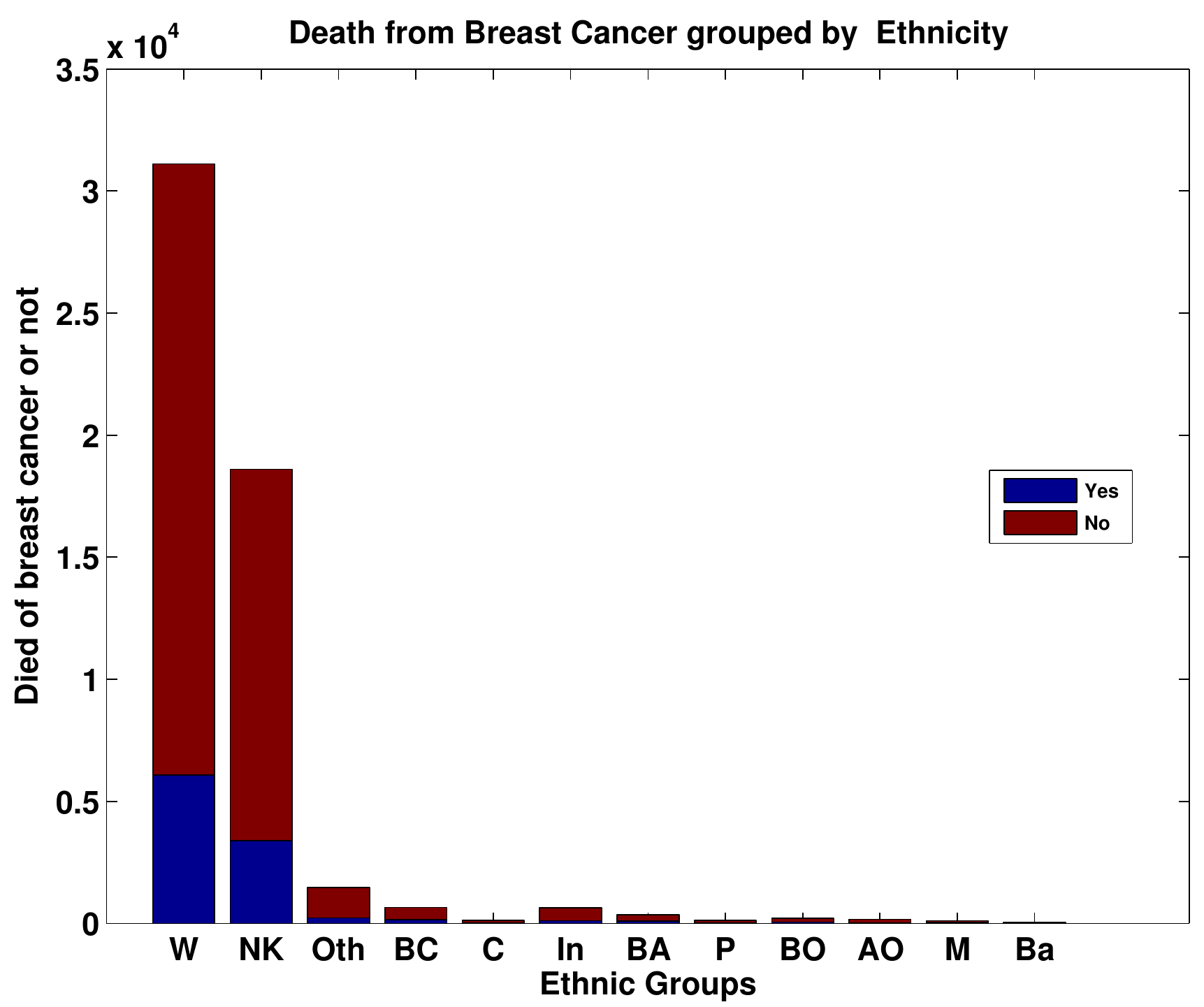}
\caption{Death and Survival from breast cancer across ethnic groups.}
\label{fig_surv_et}
\end{figure}

\begin{figure}[ht]
\centering
\includegraphics[width=3.5in]{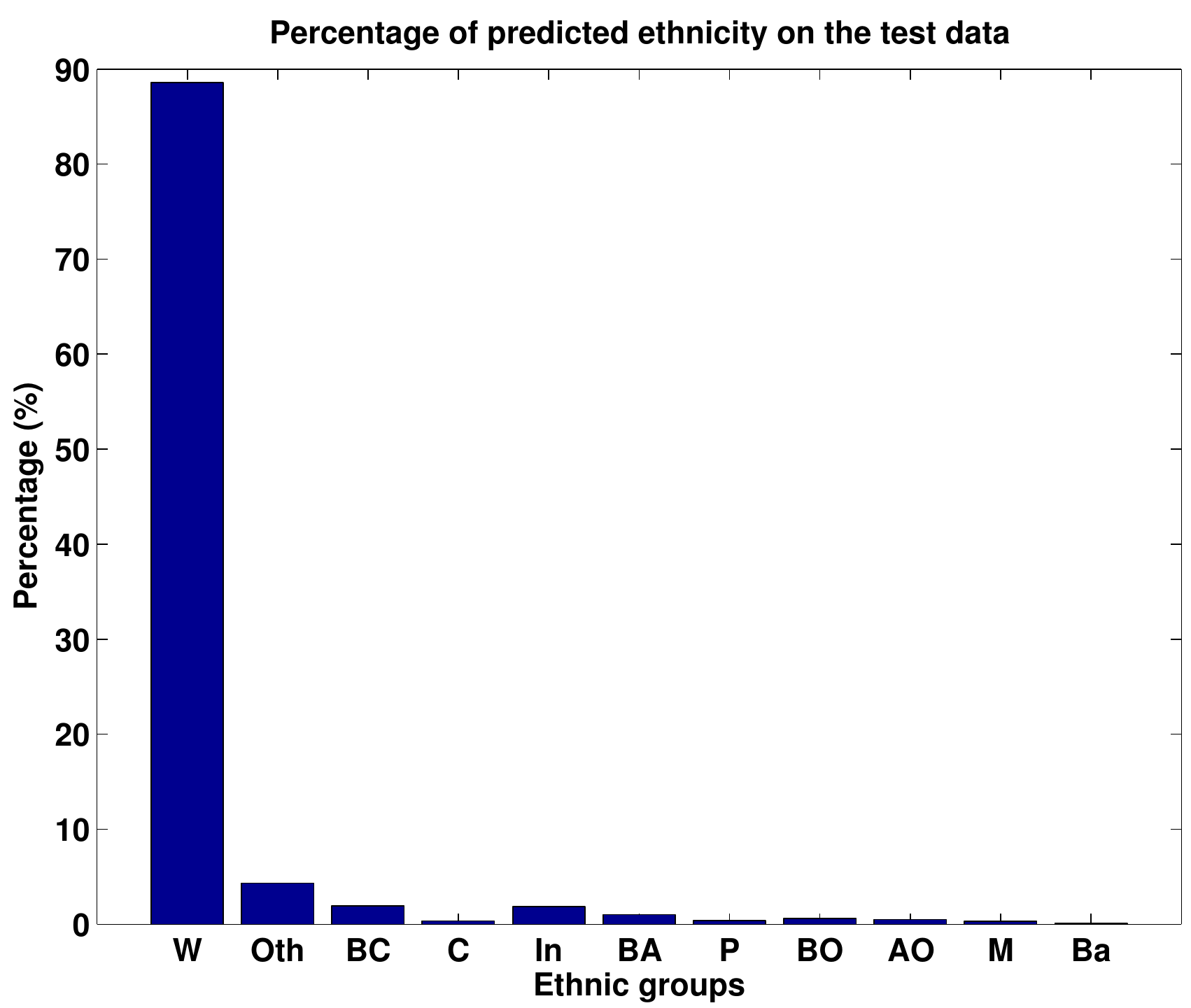}
\caption{Percentage of predicted data for each ethnicity.}
\label{fig_perc_eth}
\end{figure}

Figure~\ref{fig_perc_eth} shows the predicted values for the unknown ethnicity
records and indicates that the majority of them belong to the
white ethnic group. 

\begin{figure}[ht]
\centering
\includegraphics[width=3.5in]{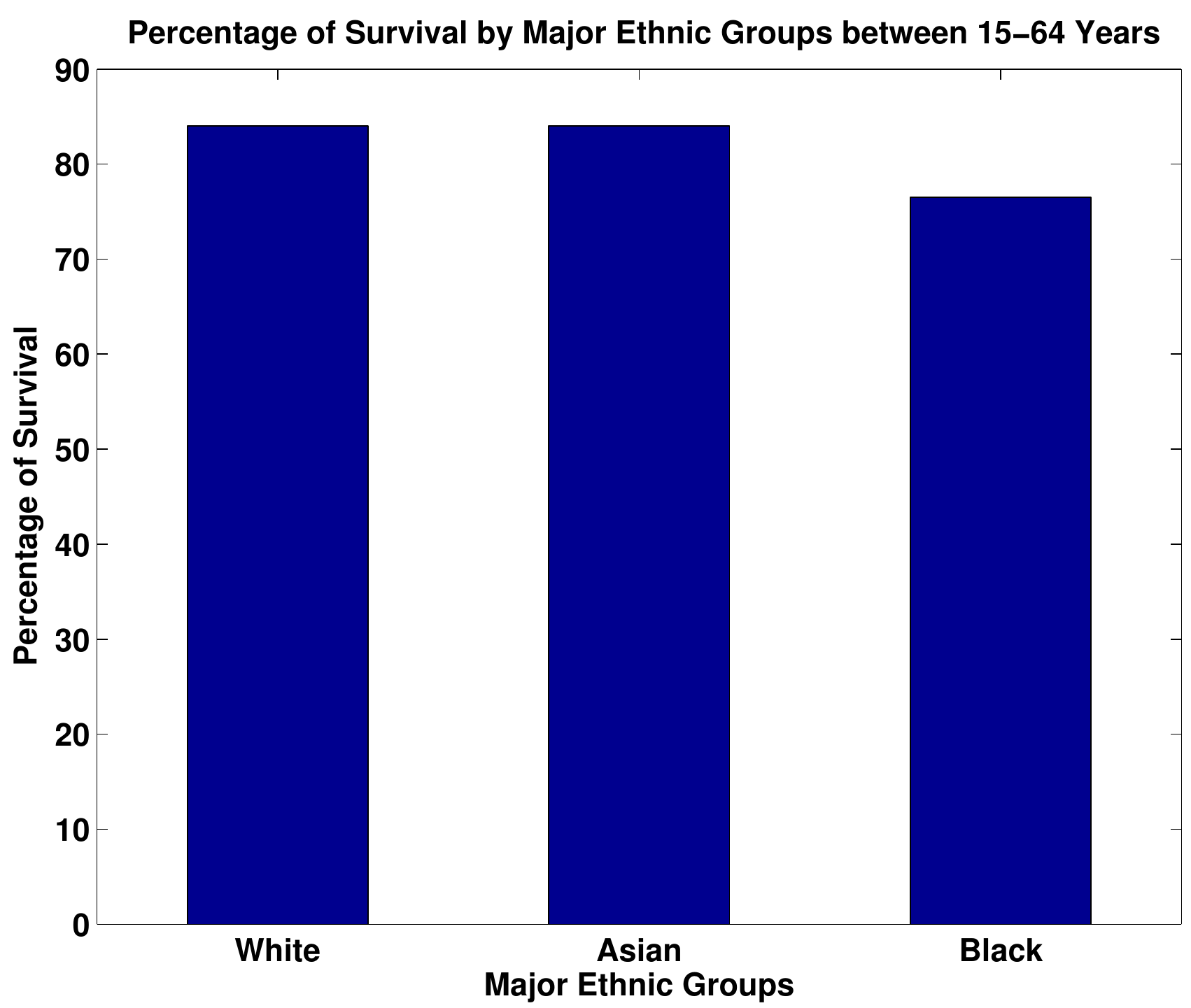}
\caption{Percentage of survival rate in white is higher than black ethnicity.}
\label{fig_surv_eth}
\end{figure}

Figure~\ref{fig_surv_eth} depicts, `For women aged between 15-64, the percentage of survival
from Breast Cancer of those of white ethnicity is likely to be
higher than those of black ethnicity' as the white women are
shown to have a 84\% survival rate compared to a 77\% survival
rate for the women belonging to the Black ethnic group.

\begin{figure}[ht]
\centering
\includegraphics[width=3.5in]{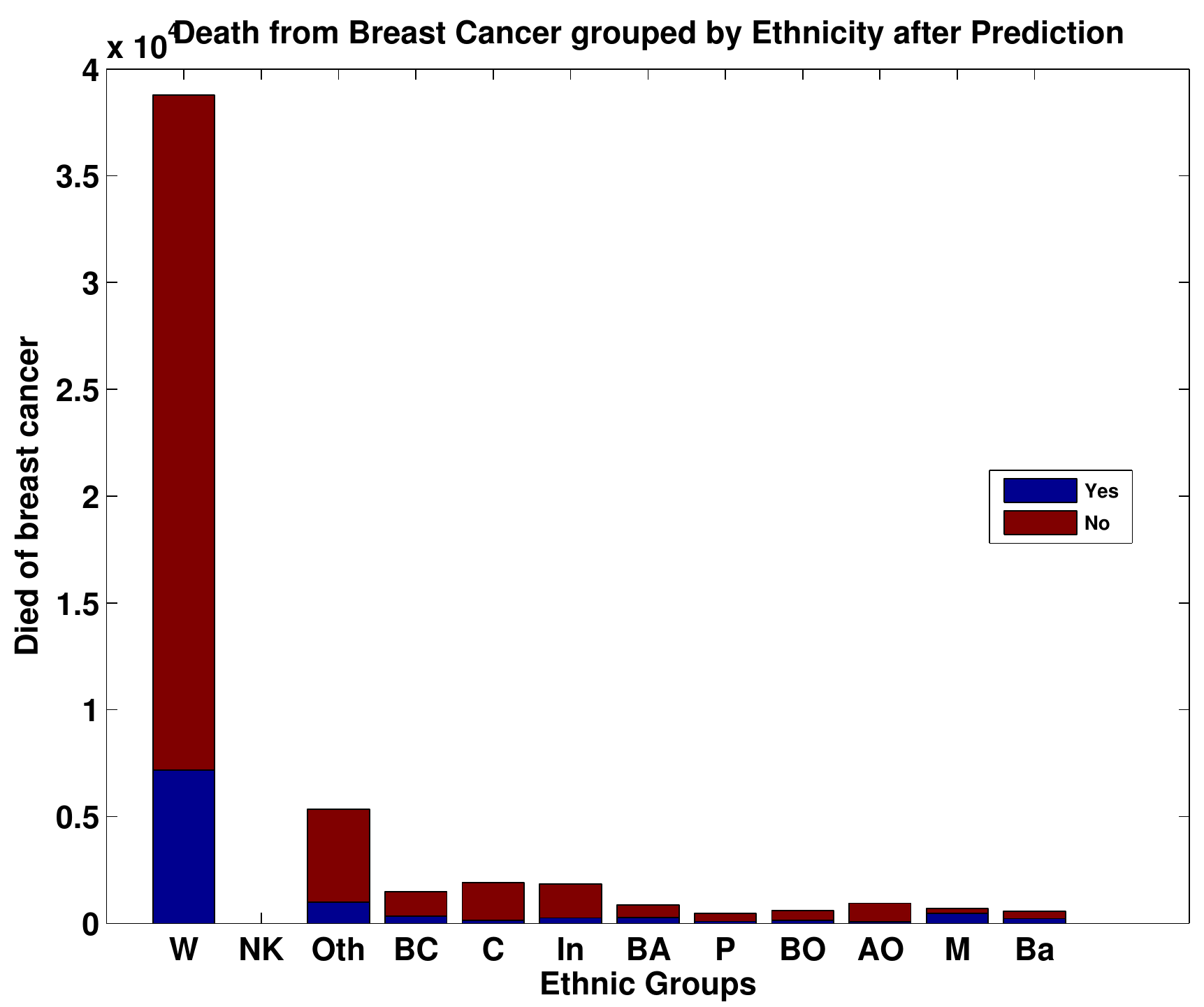}
\caption{Death from Breast Cancer grouped by Ethnicity after prediction.}
\label{fig_surv_eth_int}
\end{figure}

Figure~\ref{fig_surv_et} and Figure~\ref{fig_surv_eth_int} shows the comparison between the ethnic group
distribution before and after prediction, respectively where all
unknown records are classified into the existing ethnic groups
based on the predicted percentages obtained for each ethnicity, as shown in Figure~\ref{fig_perc_eth}. Similarly the comparison of numerals before and after imputation is shown in Table~\ref{impval}.

\subsection{Age vs survival rates}

Similarly, we produce the results for the relationship between ages and mortality. 



\begin{figure}[ht]
\centering
\includegraphics[width=3.5in]{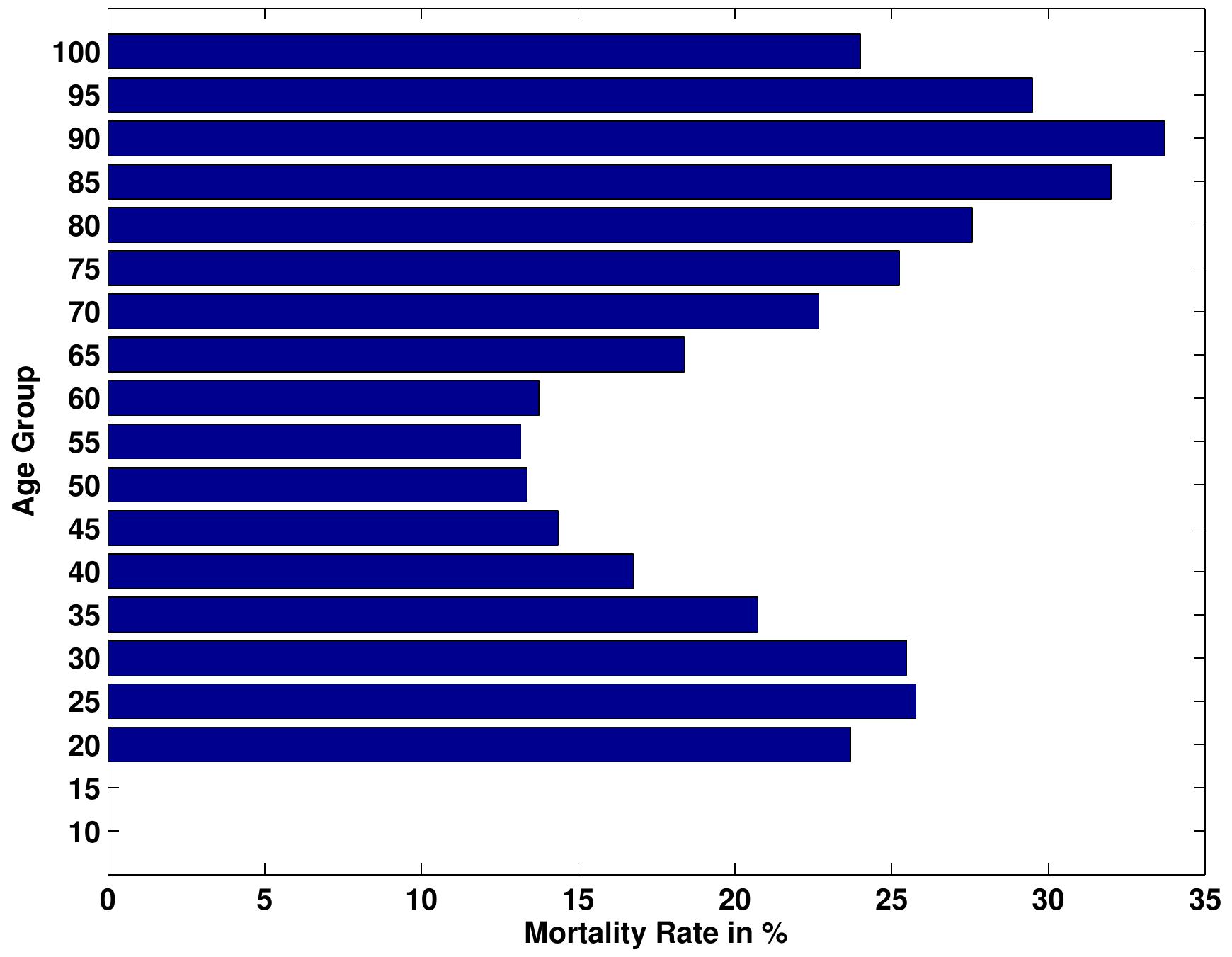}
\caption{Mortality rate according to age groups.}
\label{fig_age_mor_per}
\end{figure}

Figure~\ref{fig_age_mor_per} depicts that ages from 50 to 60 had the lowest
possibility of dying from breast cancer. The highest death
possibilities are detected in the ages between 80 and 95. Another interesting
information is that high death
possibilities are detected in the earlier ages of 20-35. This
might be because younger people are not properly informed or
do not visit their doctors in a frequent basis in comparison to
older women. 

\subsection{Income vs survival rate}

\begin{figure}[htp]
\centering
\includegraphics[width=3.5in]{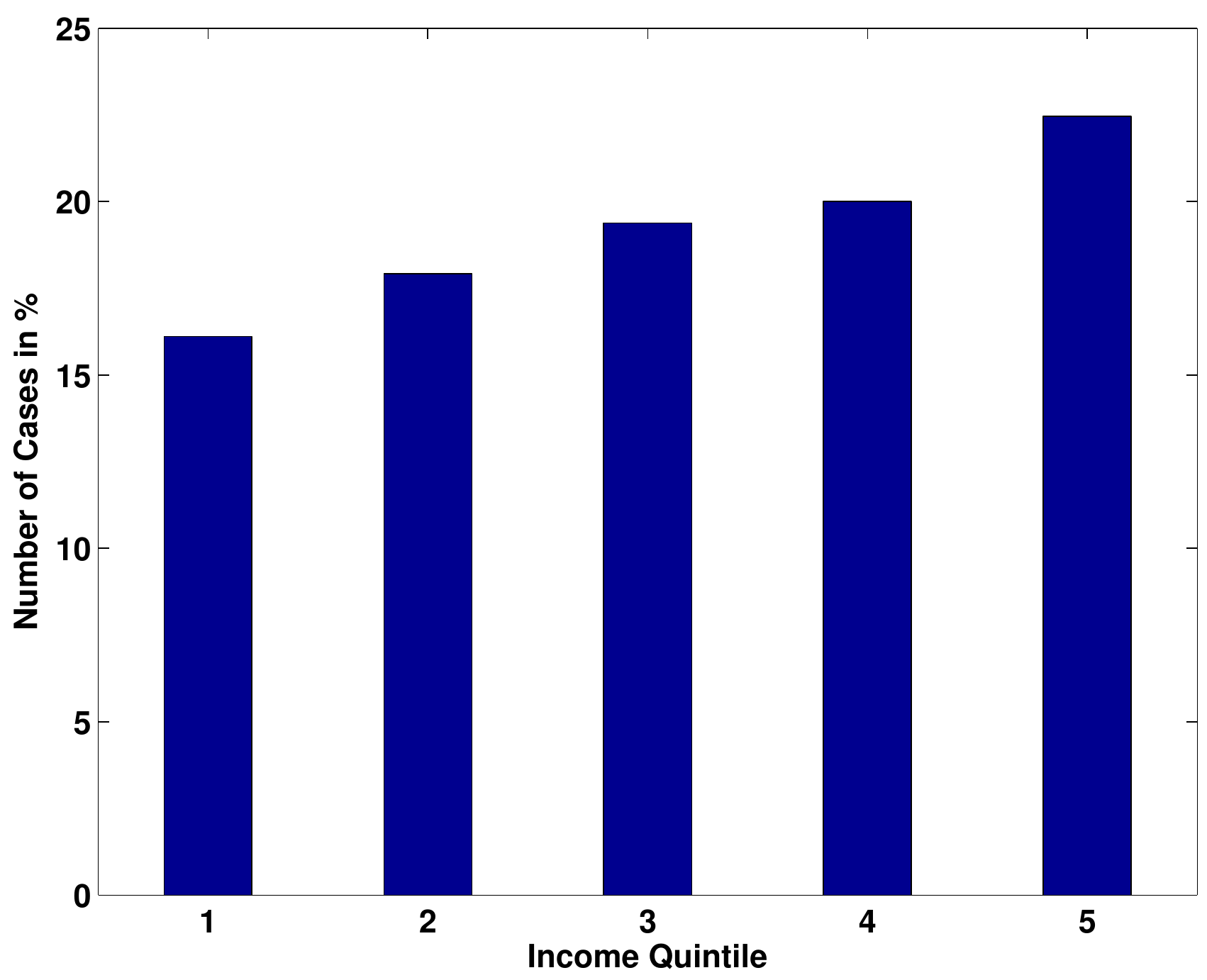}
\caption{Distribution of income within the patients.}
\label{fig_income_per}
\end{figure}
The final objective that we are interested is know whether the financial status
effects the death rate. Income feature
indicates the financial status of patients and values from
1(richest) to 5(poorer) are used. In Figure~\ref{fig_income_per}, we see that wealthier patients
have lower death rates. This is probably because they can afford
better treatment facilities.

\begin{table}[h]
\centering
\caption{comparison between the ethnic group distribution before and after imputation.}
\label{impval}
\begin{tabular}{|l|l|l|l|l|}
\hline
                   & \multicolumn{2}{l|}{\textbf{With Missing data}} & \multicolumn{2}{l|}{\textbf{After Prediction}} \\ \hline
\textbf{Ethnicity} & \textbf{Death}        & \textbf{Survival}       & \textbf{Death}       & \textbf{Survival}       \\ \hline
White              & 6072.00               & 25037.00                & 7184.00              & 31578.00                \\ \hline
Not Known          & 3385.00               & 15210.00                & 0                    & 0                       \\ \hline
Any Other          & 230.00                & 1249.00                 & 986.00               & 4358.00                 \\ \hline
Black Caribbean    & 145.00                & 507.00                  & 324.00               & 1156.00                 \\ \hline
Chinese            & 14.00                 & 103.00                  & 140.00               & 1770.00                 \\ \hline
Indian             & 113.00                & 526.00                  & 260.00               & 1576.00                 \\ \hline
Black African      & 95.00                 & 249.00                  & 274.00               & 595.00                  \\ \hline
Pakistani          & 23.00                 & 98.00                   & 92.00                & 376.00                  \\ \hline
Black Other        & 54.00                 & 158.00                  & 147.00               & 463.00                  \\ \hline
Asian Other        & 22.00                 & 148.00                  & 68.00                & 878.00                  \\ \hline
Mixed              & 30.00                 & 83.00                   & 470.00               & 234.00                  \\ \hline
Bangladeshi        & 9.00                  & 33.00                   & 223.00               & 345.00                  \\ \hline
\end{tabular}
\end{table}

\section{Conclusions and Discussions}
\label{condis}

After obtaining and appraising our results, we affirm
that the type of dataset to be classified plays a role in
selecting the appropriate distribution type
for the Bayesian classifier. Based on our results,
kernel distribution has the best F-measure percentage amongst
all other distributions.

Comparing our results with previous statistical
research in~\cite{jack2009breast} and~\cite{bradley2002race}, we can confirm that our scientific
objectives are consistent with their findings. In fact,
 referring to~\cite{rule}and~\cite{jack2009breast} approves that
white women have higher survival percentage than
black women with 91.4\% and 85\%, respectively.
Similarly, older woman and lower income groups have high mortality rates.

There always exists some limitations with the data collection. In fact, by examining the
breast cancer dataset, we can notice a clear imbalanced
number of participants between the different ethnic
groups where white females were representing more
than half of the population versus very few
numbers amongst all other ethnicity. This limitation
might have certainly misled our prediction results which
may explain the low F-measure percentages obtained on other distributions excluding kernel and Gaussian.


\section*{Acknowledgement}

The funding for this work has been provided by Department of Computing and Centre for Vision, Speech and Signal
Processing (CVSSP) - University of Surrey.



%
\bibliographystyle{IEEEtran}
\bibliography{santosh_references}

\end{document}